\documentclass[a4paper,useAMS,usenatbib]{mn2e}
\usepackage[totalwidth=480pt, totalheight=680pt]{geometry}
\usepackage{graphicx}

\title[The MicroJy and NanoJy Radio Sky: Source Population and
  Multi-wavelength Properties] {The MicroJy and NanoJy Radio Sky: Source
  Population and Multi-wavelength Properties}
\author[Paolo Padovani]{Paolo Padovani\thanks{E-mail:
ppadovan@eso.org}\\
European Southern Observatory, Karl-Schwarzschild-Str. 2,
                   D-85748 Garching bei M\"unchen, Germany\\}
\begin{document}
\newdimen\digitwidth
\setbox0=\hbox{2}
\digitwidth=\wd0
\catcode `#=\active
\def#{\kern\digitwidth}

\date{Accepted ... Received ...; in original form ...}

\pagerange{\pageref{firstpage}--\pageref{lastpage}} \pubyear{2010}

\maketitle

\label{firstpage}

\begin{abstract}
  I present simple but robust estimates of the types of sources making up
  the faint, sub-$\mu$Jy radio sky. These include, not surprisingly,
  star-forming galaxies and radio-quiet active galactic nuclei but also two
  ``new'' populations, that is low radio power ellipticals and dwarf
  galaxies, the latter likely constituting the most numerous component of
  the radio sky. I then estimate for the first time the X-ray, optical, and
  mid-infrared fluxes these objects are likely to have, which are very
  important for source identification and the synergy between the upcoming
  Square Kilometre Array (SKA) and its various pathfinders with future
  missions in other bands. On large areas of the sky the SKA, and any other
  radio telescope producing surveys down to at least the $\mu$Jy level,
  will go deeper than all currently planned (and past) sky surveys, with
  the possible exception of the optical ones from the Panoramic Survey
  Telescope and Rapid Response System and the Large Synoptic Survey
  Telescope. The Space Infrared telescope for Cosmology and Astrophysics,
  the James Webb Space Telescope (JWST), and in particular the Extremely
  Large Telescopes (ELTs) will be a match to the next generation radio
  telescopes but only on small areas and above $\sim 0.1 - 1~\mu$Jy (at 1.4
  GHz), while even the International X-ray Observatory will only be able to
  detect a small (tiny) fraction of the $\mu$Jy (nanoJy) population. On the
  other hand, most sources from currently planned all-sky surveys, with the
  likely exception of the optical ones, will have a radio counterpart
  within the reach of the SKA. JWST and the ELTs might turn out to be the
  main, or perhaps even the only, facilities capable of securing optical
  counterparts and especially redshifts of $\mu$Jy radio sources. Because
  of their sensitivity, the SKA and its pathfinders will have a huge impact
  on a number of topics in extragalactic astronomy including star-formation
  in galaxies and its co-evolution with supermassive black holes, 
  radio-loudness and radio-quietness in active galactic nuclei, 
  dwarf galaxies, and the main contributors to the radio background.
\end{abstract}

\begin{keywords}
  galaxies: active -- galaxies: dwarf -- galaxies: star formation -- radio
  continuum: general -- infrared: galaxies -- X-rays: galaxies
\end{keywords}

\section{Introduction}

The radio bright ($\ga 1$ mJy) radio sky consists for the most part of
active galactic nuclei (AGN) whose radio emission is generated from the
gravitational potential associated with a supermassive black-hole and
includes the classical extended jet and double lobe radio sources as well
as compact radio components more directly associated with the energy
generation and collimation near the central engine. Below 1 mJy there is an
increasing contribution to the radio source population from synchrotron
emission resulting from relativistic plasma ejected from supernovae
associated with massive star formation in galaxies. After years of intense
debate, however, this contribution appears not to be overwhelming, at least
down to $\sim 50~\mu$Jy. Deep ($S_{\rm 1.4GHz} \ge 42~\mu$Jy) radio
observations of the VLA-Chandra Deep Field South (CDFS), complemented by a
variety of data at other frequencies, imply a roughly 50/50 split between
star-forming galaxies (SFG) and AGN \citep{pad09}, in broad agreement with
other recent papers \citep[e.g.,][]{sey08,smo08}. About half of the AGN
are radio-quiet, that is of the type normally found in optically selected
samples and characterised by relatively low radio-to-optical flux density
ratios and radio powers \citep{pad09}. These objects represent an almost
negligible minority above 1 mJy.

This source population issue is strongly related to the very broad and
complex relationship between star formation and AGN in the Universe. At the
cosmological level, the growth of supermassive black holes in AGN appears
to be correlated with the growth of stellar mass in galaxies
\citep[e.g.,][]{mer08}. At the local level, the accreting gas feeding the
black hole at the centre of the AGN might trigger a starburst. The black
hole in turn feeds energy back to its surroundings through winds and jets,
which can compress the gas and therefore accelerate star formation but can
also blow it all away, thereby stopping accretion and star formation
altogether. Although the details are still not entirely settled, there is
however increasing evidence that in the co-evolution of supermassive black
holes and galaxies nuclear activity plays a major role through the
so-called ``AGN Feedback'' \citep[e.g.,][]{cat09}.

Radio observations afford a view of the Universe unaffected by the
absorption, which plagues most other wavelengths, and therefore provide a
vital contribution to our understanding of this co-evolution. However,
while we have a reasonably good handle on the radio evolution and
luminosity functions (LFs) of powerful sources (e.g., radio quasars), the
situation for the intrinsically fainter, and therefore more numerous, radio
sources is still murky. For example, it is still not entirely clear how
strongly low-luminosity, Fanaroff-Riley (FR) type I radio galaxies evolve
\citep{gen10}. Moreover, the widely used SFG LF of 
\cite{con89} is highly uncertain at the low end ($P_{\rm 1.4GHz} \la
10^{20}$ W Hz$^{-1}$) because, although these sources are intrinsically
very numerous, the volume within which they can be detected is small (their
``visibility function'' is very low). Indeed, a source with $P_{\rm 1.4GHz}
\sim 10^{20}$ W Hz$^{-1}$ {\it could} be detected in the deepest radio
image currently available \citep{ow08} out to $z \sim 0.05$, but it is not
\citep{str10}, because the survey area is too small. And already at $z \sim
1$, assuming a luminosity evolution $\propto (1+z)^3$, such a source would
have a flux density as low as $\sim 0.2~\mu$Jy. Finally, there are still no
published results on the radio evolution of radio-quiet AGN, which are
intrinsically weak sources ($P_{\rm 1.4GHz} \la 10^{24}$ W Hz$^{-1}$) and a
non-negligible component of the sub-mJy sky. Note that faint (sub-$\mu$Jy)
radio sources have also been proposed by \cite{sin10} as the main
contributors to the extragalactic radio background recently reported by the
ARCADE 2 collaboration \citep{fix10}. Deeper radio observations over large
areas of the sky are desperately needed to determine the LF and evolution
of the most common radio sources in the Universe.

These will soon be realised, as radio astronomy is at the verge of a
revolution, which will usher in an era of large area surveys reaching flux
density limits well below current ones. The Square Kilometre Array
(SKA)\footnote{http://www.skatelescope.org}, in fact, will offer an
observing window between 70 MHz and 10 GHz extending well into the {\it
  nanoJy} regime with unprecedented versatility.
The field of view will be large, up to $\sim 200$ deg$^2$ below 0.3 GHz and
possibly reaching $\sim 25$ deg$^2$ at 1.4 GHz. First science with $\sim
10\%$ SKA should be near the end of this decade. Location will be in the
southern hemisphere, either Australia or South Africa. Many surveys are
being planned with the SKA, possibly comprising an ``all-sky" 1 $\mu$Jy
survey at 1.4 GHz and an HI survey out to redshift $\sim 1.5$, which should
include $\sim 10^9$ galaxies.

The SKA will not be the only participant to this revolution. The LOw
Frequency ARray
(LOFAR)\footnote{http://www.astron.nl/radio-observatory/as\-tronomers/lofar-astronomers}
has recently started operations and will carry out large area surveys at
15, 30, 60, 120 and 200 MHz \citep[see][for details]{mor09}, opening up a
whole new region of parameter space at low radio frequencies. Many other
radio telescopes are currently under construction in the lead-up to the SKA
including the Expanded Very Large Array
(EVLA)\footnote{http://science.nrao.edu/evla}, the Australian Square
Kilometre Array Pathfinder
(ASKAP)\footnote{http://www.atnf.csiro.au/projects/askap/}, the Allen
Telescope Array\footnote{http://ral.berkeley.edu/ata},
Apertif\footnote{http://www.astron.nl/general/apertif/apertif}, and
Meerkat\footnote{http://www.ska.ac.za/meerkat}.  These projects will survey
the sky vastly faster than is possible with existing radio telescopes
producing surveys covering large areas of the sky down to fainter flux
densities than presently available.

Current deep surveys include a number of VLA small area surveys below 0.1
mJy at a few GHz, reaching a maximum area of $\sim 2$ deg$^2$
\citep[VLA-COSMOS;][]{bon08} and a minimum flux density $\sim 15~\mu$Jy at
1.4 GHz \citep[SWIRE;][]{ow08} and $\sim 7.5~\mu$Jy at 8.4 GHz \citep[SA
13;][]{ fom02}. The NRAO VLA Sky Survey \citep[NVSS;][]{con98} and the
Faint Images of the Radio Sky at Twenty centimeters \citep[FIRST;][]{bec95}
only reach 1.4 GHz limits of 2.5 and 1 mJy respectively but cover much
larger areas (82\% and 22\% of the sky respectively).

What lies beneath the surface of the deepest surveys we currently have?
Predictions for the source population at radio flux densities $< 1~\mu$Jy
have been made, amongst others, by \cite{hop00}, \cite{win03},
\cite{jac04}, \cite{jr04}, and \cite{wil08}. These papers have presented
detailed estimates for the number counts of faint radio sources predicting,
for example, that SFG should make up $\sim 90\%$ of the total population at
$S_{\rm 1.4GHz} \sim 1~\mu$Jy \citep{wil08} but had to rely, for obvious
reasons, on extrapolations.
Only the last two papers include radio-quiet AGN in their modelling by
converting their X-ray LF to the radio band assuming a linear correlation
between radio and X-ray powers. Most importantly, as described below, two
crucial constituents of the sub-$\mu$Jy sky have been excluded by all of
these studies.

The first aim of this paper is then to have a broad look at the likely
astrophysical populations, which make up the very faint radio sky. But
detecting sources is only part of the story, as then comes the
identification process. This requires a wealth of multi-wavelength data,
ranging from the optical/near-IR imaging needed to provide an optical
counterpart and, when needed, photometric redshifts, to the optical/near-IR
spectra required to estimate a redshift, and hence the distance of sources,
to the X-ray data, which are vital to separate AGN from SFG \citep[e.g.,]
[and references therein]{pad09}, to the mid-infrared colours, which provide
additional information on this separation \citep[e.g.,][] {saj05}. The
second goal of this work is then to provide estimates for the the X-ray,
optical, and mid-infrared fluxes these sources are likely to have. Optical
magnitudes will also determine how feasible it will be to obtain redshifts
for them. This kind of information is important for planning purposes, to
be ready to take full advantage of the new, deep radio data, and also to
maximise the synergy between the SKA and its pathfinders, and present but
also, most importantly, future missions. To the best of my knowledge, these
estimates have never been made before.

Throughout this paper spectral indices are written $S_{\nu} \propto
\nu^{-\alpha}$, magnitudes are in the AB system, and the values $H_0 = 70$
km s$^{-1}$ Mpc$^{-1}$, $\Omega_{\rm M} = 0.3$, and $\Omega_{\rm \Lambda} =
0.7$ have been used, unless otherwise noted. 
All the radio results refer to 1.4 GHz. Although I
mention only the SKA throughout the paper, my results obviously apply to
any other radio telescope producing surveys reaching the $\mu$Jy level.  I
also make reference to a large number of missions and projects, for which I
give mainly sensitivity information. Readers wanting to know more should
consult the relevant World Wide Web pages, whose addresses are provided in
the text.

\section{MicroJy and nanoJy radio source population}\label{pop}

I present here a simple approach to study the radio sky source population,
based on only two parameters: the smallest flux density and the largest
surface density of radio sources. The main idea is to provide robust
results based on some basic observables and to pay particular attention to
{\it all} populations reaching below the $\mu$Jy level.

\begin{figure*}
\centering
\includegraphics[width=16.0cm]{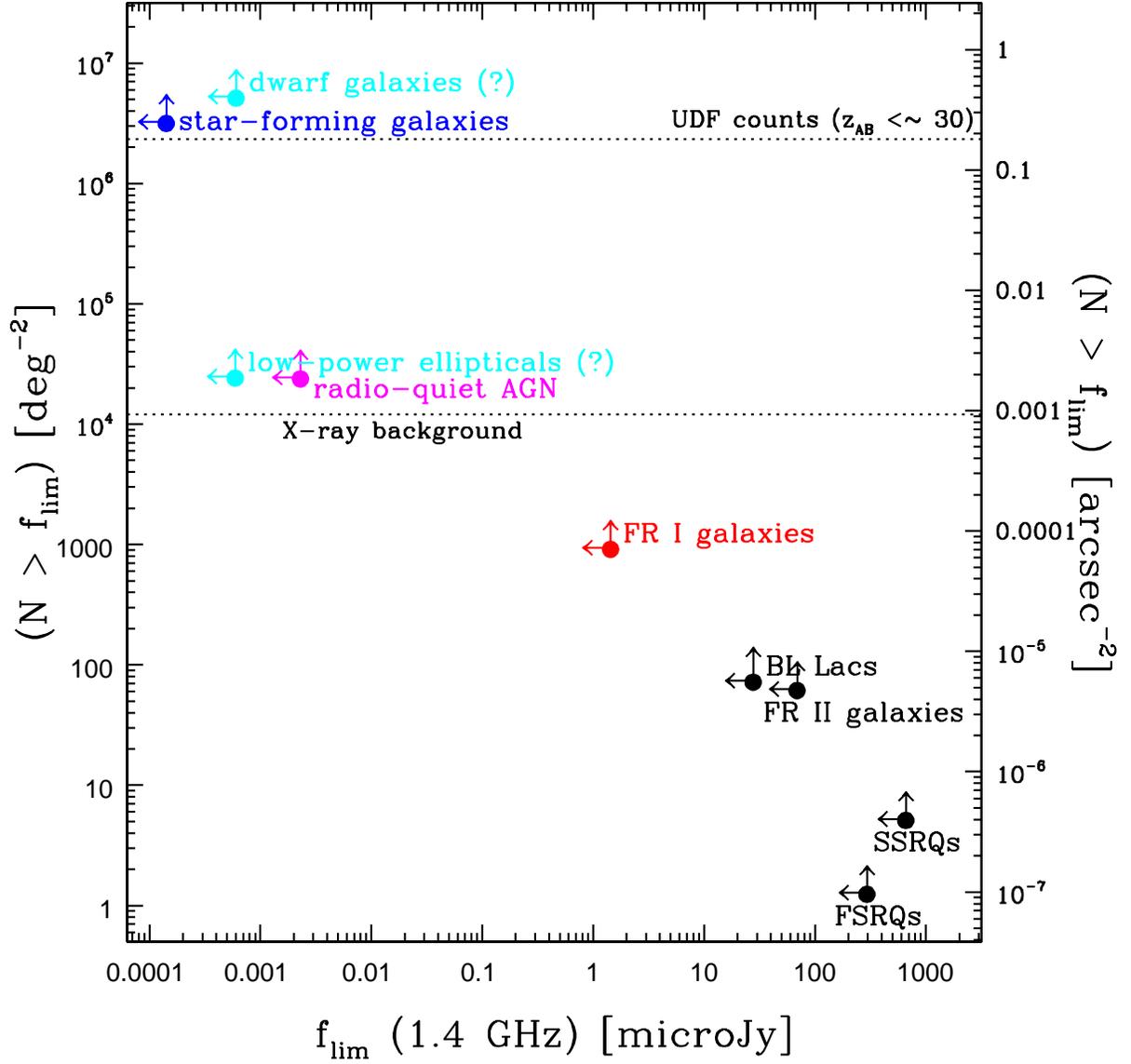}
\caption{The largest surface density vs. the smallest flux density for
  various classes of radio sources. The two horizontal lines denote, from
  top to bottom, the surface density of the optical sources in the Hubble
  Ultra Deep Field and the surface density of the AGN needed to explain the
  X-ray background. See text for more details.}
\label{flim} 
\end{figure*}

The smallest flux density $f_{\rm lim}$ of a population of sources depends
on the minimum radio power at $z \sim 0$, $P_{\rm min}(0)$, the maximum
redshift of the sources, $z_{\rm max}$, and any luminosity evolution
$le(z)$, where $P(z) = P(0) \times le(z)$. If evolution peaks at $z_{\rm
  top}~(\le z_{\rm max})$ and then stops, then
\begin{equation}
f_{\rm lim} = {P_{\rm min}(0) le(z_{\rm top}) (1+z_{\rm
    max})^{1-\alpha}\over 4 \pi D_L^2(z_{\rm max})}
\label{eq_flim}
\end{equation}
where $D_L(z)$ is the luminosity distance. Luminosity evolution makes
sources brighter and therefore increases $f_{\rm lim}$.

The largest surface density of a population, $N(> f_{\rm lim})$, depends on
its number density at $z \sim 0$, $N_{\rm T}(0)$, the maximum redshift
$z_{\rm max}$, and any density evolution $de(z)$, where $N_{\rm T}(z) =
N_{\rm T}(0) \times de(z)$. Then

\begin{equation}
N(> f_{\rm lim}) = {N_{\rm T}(0) \over 4 \pi} \int_0^{z_{\rm max}} de(z)
dV/dz~~~~{\rm sr^{-1}}
\label{eq_N}
\end{equation}
where $dV/dz$ is the derivative of the comoving volume. In case of no
density evolution $N(> f_{\rm lim}) = N_{\rm T}(0) V(z_{\rm
  max})/4 \pi~{\rm sr^{-1}}$. 
  
Information on the local LF, needed to derive $N_{\rm T}(0)$ and $P_{\rm
  min}(0)$, and the evolution for various classes was derived from a
variety of sources. Namely:

\begin{enumerate}

\item the LF for SFG is that of \cite{sad02}, which at the low end is
  complemented by that of \cite{con89}; luminosity and density evolution
  and $z_{\rm top}$ are from \cite{hop04};

\item the LF for radio-quiet AGN is built using data from \cite{ru96}
  (12$\mu$m and CfA Seyfert samples) and Padovani et al. (in preparation:
  VLA-CDFS sample). Evolution is from Padovani et al. (in preparation), 
  while $z_{\rm top}$ is from the X-ray band \citep{ha05};

\item the LF for FR Is derives from that of \cite{up95}, which agrees well
  with the very recent derivation of \cite{gen10} apart from the first bin;
  the LF was then modified accordingly and the values of $N_{\rm T}(0)$ and
  $P_{\rm min}(0)$ were derived. \cite{gen10} find evidence of evolution
  for FR Is but only at $P_{\rm 1.4GHz} \ga 10^{25}$ W Hz$^{-1}$. Since the
  total surface density in eq. \ref{eq_N} depends on the most numerous, and
  therefore least luminous, sources (which reach $P_{\rm 1.4GHz} \sim
  10^{23}$ W Hz$^{-1}$) no evolution was assumed;

\item the LF for FR IIs is also based on that of \cite{up95}, which agrees
  quite well with that of \cite{gen10} down to $P_{\rm 1.4GHz} \approx
  10^{24}$ W Hz$^{-1}$. Below this value the LF is basically
  undetermined so no modification was done. Evolution was taken from
  \cite{up95};

\item the LF for flat-spectrum radio quasars (FSRQs), steep-spectrum radio
  quasars (SSRQs), and BL Lacs are based on those of \cite{up95}, which
  were calculated from those of FR IIs and FR Is respectively, based on a
  beaming model. These have been shown to agree with LFs derived from
  recent samples \citep{pad07}. Since the FR I LF was modified, the beamed
  LF for BL Lacs was also changed accordingly at low powers before deriving
  $N_{\rm T}(0)$ and $P_{\rm min}(0)$. Evolutionary parameters are from
  \cite{up95}, while $z_{\rm top}$ values for FSRQs, SSRQs, and FR
  IIs\footnote{According to unified schemes \citep[see, e.g.,][]{up95},
    SSRQs and FR IIs, being FSRQs seen at larger angles with respect to
    the line of sight, need to share the same $z_{\rm top}$.} come from
  \cite{dez05}.

\end{enumerate}

Values of $z_{\rm max}$ were fixed to the highest redshift of the class
under consideration, that is $\sim$ 6, 6.5, and 5.5 for SFG, radio-quiet
AGN, and FSRQs, SSRQs, and FR IIs respectively, apart from FR Is and BL
Lacs, for which $z_{\rm max} = 3$ was assumed. Finally, the blazar
catalogue of \cite{mas09} and the AGN catalogue of \cite{pad97} were
checked to see if any sources had radio power below the adopted
$P_{\rm min}$ values. None was found.

\begin{table*}
\caption{Radio populations parameters}
\begin{tabular}{lrllcll}
  Class & $N_{\rm T}(0)$ & $##P_{\rm min}(0)$& LE & DE & $z_{\rm top}$ & $z_{\rm max}$\\
                &  Gpc$^{-3}$ & ##W Hz$^{-1}$  &  & & &  \\
\hline
 FSRQs &12&$##2\times {10^{24}}$&$\exp[T(z)/0.23]^a$& ... & 2.25 & 5.5\\
 SSRQs &59&$##3\times {10^{24}}$&$\exp[T(z)/0.15]^a$& ... & 2.25 & 5.5\\
 FR IIs &590&$##3\times {10^{24}}$&$\exp[T(z)/0.26]^a$& ... & 2.25 & 5.5\\
 BL Lacs &2,310&$##{10^{23}}$&$\exp[T(z)/0.32]^a$& ... & ... & 3.0\\
 FR Is &29,300&$##{10^{23}}$&...& ... & ... & 3.0\\
 RQ AGN & $3.9\times 10^5$& $##5\times 10^{19}$& $(1+z)^{2.4}$& ... & 1.7& 6.5 \\
 SFGs & $4.5\times 10^7$& $##2\times 10^{18}$& $(1+z)^{2.7}$& $(1+z)^{0.15}$ &
2.0& 6.0 \\
 Dwarf Galaxies & $2.0\times 10^8$& $<2\times 10^{18}$& $(1+z)^{2.7}$& ... &
 2.0& 3.0 \\
 Low-power Ellipticals & $4.8\times 10^6$& $<3\times 10^{19}$& ... &
$(1+z)^{-1.7}$ & ... & 3.0 \\ \hline
\multicolumn{5}{l}{\footnotesize $^a$$H_0 = 50$, $q_0=0$; $T(z)$ is the
  look-back time}\\
\end{tabular}
\label{tab:pop}
\end{table*}
 
  Table \ref{tab:pop} summarises the parameters used in eqs. \ref{eq_flim}
  and \ref{eq_N}. $N_{\rm T}(0)$ values were derived in most cases from 
  simple fits to the LF and should be considered approximate. 
  The $N_{\rm T}(0)$ and $P_{\rm min}(0)$ values from
  \cite{up95} have been converted to $H_0 = 70$ from $H_0 = 50$ while the
  evolutionary parameters still refer to an $H_0 = 50$, $q_0=0$ cosmology
  (as a simple conversion in this case is not possible). Note that,
  although more complex evolutionary models than those used here have been
  adopted in the literature for ``classical'' powerful radio sources, the
  consensus is that such sources reach only the $\approx$ mJy level
  \citep[e.g.,][]{jac04,wil08}. This is in agreement with my own results
  below and in any case well above the flux densities of interest here.

The resulting flux and surface density limits are shown in Fig. \ref{flim}.
These values should be considered as robust upper and lower limits
respectively, because: 1. one cannot exclude that lower-power, and
therefore more numerous, objects exist; 2. $z_{\rm max}$ could be larger
than assumed. The latter parameter has a stronger influence on flux than on
surface density. For example, for $z_{\rm max} = 6$, eq. \ref{eq_flim}
shows that the former value decreases by a factor $(4/7)^{1-\alpha_r}
[D_L(6)/D_L(3)]^2 \sim 4.4$ ($\alpha_r = 0.7$) as compared to $z_{\rm max}
= 3$, while from eq. \ref{eq_N} and in the case of no density evolution the
latter increases by a factor $V(6)/V(3) \sim 2.2$. If $z_{\rm max}$
increases from 6 to 10, the two values are instead $\sim 2.8$ and $\sim
1.5$.

Fig. \ref{flim} shows that the most powerful radio sources, that is FRSQs,
SSRQs, and FR IIs are, not surprisingly, the ones having the largest flux
density ($\approx 0.1 - 1$ mJy) and the smallest surface density ($\approx
1 - 50~{\rm deg}^{-2}$) limits. BL Lacs are only slightly fainter then FR
IIs, while FR Is are the only radio-loud sources reaching $\approx
1~\mu$Jy. Radio-quiet AGN and SFG are the faintest classes, going into the
nanoJy regime, with SFG dominating the faint radio sky (amongst
``classical" radio sources: see below). Indeed, the differential counts of
\cite{wil08} predict a strong dominance ($\sim 90\%$) of SFG at $1~\mu$Jy,
which gets slightly weaker at nanoJy flux densities.

In Fig. \ref{flim} I have also plotted two horizontal lines. The one at $N
\approx 10^4$ deg$^{-2}$ denotes the surface density of the AGN needed to
explain the X-ray background \citep[Gilli, private communication, based
on][]{gil07}. Since these for the most part are radio-quiet AGN, its
proximity to the estimated surface density of this class is reassuring. The
other line at $N \approx 2 \times 10^6$ deg$^{-2}$ indicates the surface
density of the optical sources in the Hubble Ultra Deep Field (UDF)
\citep{bec06}, thought to be mostly star-forming galaxies. The fact that
this is very close to the completely independent estimate for SFG in the
radio band shows that the latter is perfectly plausible.

I argue that two other populations play a major role at $S_{\rm 1.4GHz} <
1~\mu$Jy. The first one is that of low-power ellipticals. It has been know
for quite some time that ellipticals of similar optical luminosity vary
widely in radio power, with some (non-dwarf) galaxies having $P_{\rm 5GHz}
< 2 \times 10^{19}$ W Hz$^{-1}$ \citep[e.g.,][converting from their
cosmology]{sad89}. More recently, \cite{cap09} have shown that 82\% of
early-type galaxies in the Virgo cluster with $B_{\rm T} < 14.4$ are
undetected at a flux density limit of $\sim 0.1$ Jy, which implies core
radio powers $P_{\rm 8.4GHz} < 4 \times 10^{18}$ W Hz$^{-1}$. \cite{mil09}
have studied the radio LF in the Coma cluster and found that 58\% of red
sequence galaxies with $M_r \le -20.5$ are undetected at about $28~\mu$Jy
r.m.s. Stacking these sources, they obtained a detection corresponding to
$P_{\rm 1.4GHz} \sim 3 \times 10^{19}$ W Hz$^{-1}$. These faint ellipticals
are {\it not} represented in previous models of the sub-$\mu$Jy sky: for
example, the lower limit of the radio-loud AGN LF in \cite{wil08} is
$P_{\rm 1.4GHz} = 2 \times 10^{20}$ W Hz$^{-1}$. We have no information on
the evolution of these radio sources but, based on the results of low-power
($P_{\rm 1.4GHz} < 10^{25}$ W Hz$^{-1}$) radio galaxies \citep[e.g.,][and
references therein]{gen10}, it seems plausible to assume no luminosity
evolution. Taking $P_{\rm min}  <  3 \times 10^{19}$ W Hz$^{-1}$ and $z_{\rm
  max} = 3$ one derives $f_{\rm lim} < 0.6$ nanoJy (assuming the same
luminosity evolution as found by \cite{cim06} for ellipticals in the B-band
one would instead get $f_{\rm lim} < 5$ nanoJy).  
(Note that since we only have an 
upper limit on $P_{\rm min}$ in this case $f_{\rm lim}$ is an even more robust
upper limit than for the previously discussed classes.)

As regards their surface density, I have taken the number density of all
early-type galaxies from the LF in \cite{delap03} (the two-wing Gaussian
fit for $R_c \le 21.5$ in their Tab. 7), which agrees within $\sim 20\%$
with the Schechter fit to the local Sloan Digital Sky Survey (SDSS) LF done
by \cite{bel04}. \cite{kri08} have recently suggested a decrease of a
factor $\sim 8$ in the number density of high-mass ($> 10^{11} M_{\odot}$)
early-type galaxies between $z \sim 0$ and 2.3. \cite{cim06} have found a
similar trend for low-mass ($< 10^{11} M_{\odot}$) early-type galaxies
between $z \sim 0$ and 1.2. Assuming that all ellipticals are radio sources
at some (very low)
level, such a density evolution, and $z_{\rm max} = 3$, I get from
eq. \ref{eq_N} a limiting surface density for low-power ellipticals
$\approx 2.4 \times 10^4$ deg$^{-2}$, of the same order as that of
radio-quiet AGN.

The other population missing from previous studies is that of dwarf
galaxies, which are very faint and constitute the most numerous
extragalactic population. This class includes dwarf spheroidals and
ellipticals, dwarf irregulars, and blue compact dwarf galaxies (BCDs), and
it has never been considered for the simple reason that its radio LF and
evolution has never been determined. But the simple approach adopted here
can provide us with some idea of how faint and how numerous these sources
are going to be in the sub-$\mu$Jy sky.

The determination of the LF of dwarf galaxies is not an easy task as it
requires a thorough understanding of selection effects, due to their low
surface brightness. The LF appears also to be environment-dependent
\citep[e.g.,][]{fer91}. I have taken the SDSS LF of \cite{bla05}, which is
corrected for surface brightness incompleteness, and derived the number
density of dwarf galaxies (defined by $M_{\rm r} \la -18.8$ [equivalent to
$M_{\rm B} \la -18.1$], which is where there is an upturn in the slope of
the LF). Available data are consistent with no density evolution at the
faint end of the LF up to at least $z \approx 3$
\citep[e.g.,][]{coo05,sal08}. With these assumptions I derive from
eq. \ref{eq_N} a (likely) conservative limiting surface density $\approx 5
\times 10^6$ deg$^{-2}$, higher than all other classes.


As regards flux density, \cite{le05} have shown that $P_{\rm min,1.4GHz}$
for dwarf galaxies is $< 1.6 \times 10^{18}$ W Hz$^{-1}$. Since most
galaxies at the faint end of the LF are blue \citep[e.g.,][]{bla05,sal08},
most dwarfs in the Universe should be of the star-forming type and I then
assume the same luminosity evolution as for SFG. For $z_{\rm max} = 3$ I
then get $f_{\rm lim} < 0.6$ nanoJy (no evolution would imply $f_{\rm
  lim} < 0.03$ nanoJy).  
(As was the case for low-power ellipticals, since we only have an 
upper limit on $P_{\rm min}$ $f_{\rm lim}$ is an extremely robust upper limit.)
In summary, {\it dwarf galaxies are likely to be
  the most numerous component of the faint radio sky}.

Although this paper deals with extragalactic sources, one might worry that
stellar objects could also contribute substantially to the $\mu$Jy and
nanoJy sky. This is extremely unlikely. The radio thermal component of the
Sun at the distance of $\alpha$ Centauri would have a flux density $\sim 5
- 30~\mu$Jy \citep{whi04} (where the lower value refers to the quiet state
and the higher one to the flaring one), which means $\sim 0.7 - 4$ nanoJy
at 10 Kpc. Moreover, a Sun-like star at 10 Kpc would have a non-thermal
flux density $\sim 0.001$ nanoJy \citep{sea97}. Given that the most common
main-sequence stars are of the M type, which are more than one order of
magnitude less luminous than the Sun, the bulk of stellar radio emitters
will be very faint ($< 1$ nanoJy).

\section{Multi-wavelength properties of microJy and nanoJy radio sources}\label{multi}

I estimate here the X-ray, optical, and mid-infrared fluxes radio-quiet
AGN, SFG, and FR Is should have at the $\mu$Jy and nanoJy flux density
levels. Low-power ellipticals and dwarf galaxies require some discussion.

\cite{bal06} have studied the multi-wavelength characteristics of very low
radio power ellipticals with $10^{19} < P_{\rm 5GHz} < 3 \times 10^{24}$ W
Hz$^{-1}$ ($\langle P_{\rm 5GHz} \rangle \sim 6 \times 10^{21}$ W
Hz$^{-1}$). These sources can be considered as miniature radio-galaxies, in
the sense that their {\it nuclear} properties are scaled down versions of
those of low-luminosity, FR I radio galaxies.  
Indeed, if one separates radio galaxies on the basis of their nuclear
activity into high-excitation (HERGs) and low-excitation (LERGs)
radio-galaxies, almost all FR Is are LERGs and most FR IIs are HERGs,
although there is a population of FR II LERGs as well
\citep[e.g.,][]{la94}. In this scheme, low-power ellipticals could be
considered as the natural extension of LERGs to lower radio luminosities.
In the following I will then
assume that low-power ellipticals have the same multi-wavelength properties
as their higher-power relatives, with the obvious caveat that, for the same
galaxy optical magnitude they will have a much lower radio emission, and
therefore will be characterised by a much lower radio-to-optical flux
density ratio. 
The situation for dwarf galaxies is more complex. In dwarf spheroidals and
ellipticals very little, if any, star formation is going on now, although
their star formation histories are complex \citep[e.g.,][]{fer94}. Dwarf
irregulars appear to be low mass versions of large spirals
\citep[e.g.,][]{kle86,le05}, while BCDs fall, on average, at the high star
formation rate end \citep{hun04}.

Since dwarf galaxies are intrinsically weak, our knowledge of their
spectral energy distributions (SEDs) is quite scanty and based on small
samples, which are most likely affected by selection effects. My working
hypothesis, which is not contradicted by the data (see below), will be that
dwarf spheroidals and ellipticals have SEDs similar to those of low-power
ellipticals, while dwarf irregulars and BCDs are mini-spirals. As mentioned
above, the faint radio sky should be dominated by the star-forming type of
dwarfs.

In the following the multi-wavelength fluxes of the various classes of
faint radio sources are estimated using three (two for the near-IR band)
different methods. Readers not interested in the details should skip
directly to Sect. \ref{x-ray_fluxes}, \ref{opt_mag}, and \ref{IR_fluxes}.

\subsection{X-ray Band}

\subsubsection{Typical X-ray-to-radio flux density ratios}\label{ratios}

\cite{ran03} have shown that X-ray and radio powers in SFG are strongly
correlated, likely because they both trace the star formation rate. Their
typical X-ray-to-radio flux ratio is $\langle \log(f_{\rm 0.5 - 2
  keV}/S_{1.4GHz}) \rangle = -17.9\pm0.3$ (where the X-ray flux is in
c.g.s. units and the radio flux density is in $\mu$Jy). Although derived
for local galaxies, the correlation appears to hold at higher redshifts as
well \citep[e.g.,][]{ran03,pad09}.

\cite{ott05} have studied the X-ray properties of eight dwarf galaxies
undergoing starburst observed with {\it Chandra}. I derive $\langle
\log(f_{\rm 0.5 - 2 keV}/S_{1.4GHz}) \rangle \sim -18$ (where I have used
observed X-ray fluxes converted to the $0.5-2$ keV band, taking into
account both extended and point source emission, and radio data from the
literature), perfectly consistent with the mean value for SFG.

X-ray and radio powers are also correlated in radio-quiet AGN, with a
relatively large dispersion \citep[see, e.g., Fig. 13 of][]{bri00}. In
deriving typical X-ray-to-radio flux density ratios for this class, it is
important to keep in mind that, as discussed in \cite{pad09}, X-ray
selection will tend to favour sources with relatively large X-ray-to-radio
flux ratios while the opposite will be true for radio selection. I have
used the VLA-CDFS radio-quiet AGN sample selected by \cite{pad09} (and
refined by Padovani et al. in preparation) and the X-ray fluxes provided by
\cite{toz09} to obtain a K-corrected value of $\langle \log(f_{\rm 0.5 - 2
  keV}/S_{1.4GHz}) \rangle = -17.2\pm0.9$ for radio-selected, radio-quiet
AGN. (For $\alpha_x \approx 0.9 - 1.1$, the range covered by (unabsorbed) 
RQ AGN and spirals, and synchrotron emission ($\alpha_r \sim 0.7$), 
K-correction effects almost cancel out.)
Note that these radio-quiet AGN include both broad-lined (type 1) and
narrow-lined (type 2) AGN. Whenever possible, X-ray fluxes were corrected
for absorption \citep[see][for details]{toz09}. As regards X-ray selected,
radio quiet AGN, I use the results of \cite{pad09} on the hard X-ray
selected sample of \cite{pol07} ($f_{\rm 2-10 keV} > 10^{-14}$ erg
cm$^{-2}$ s$^{-1}$) derived using survival analysis due to the many upper
limits on radio flux densities. By converting them to the $0.5 - 2$ keV
band I obtain a K-corrected mean value $\langle \log(f_{\rm 0.5 - 2
  keV}/S_{1.4GHz}) \rangle \sim -14.8$.

X-ray and radio emission correlate in radio galaxies as well, albeit with a
large scatter. Fig. 5 of \cite{pad09} shows that radio galaxies have an
X-ray-to-radio power ratio of the same order of that of SFG but smaller
than that of quasars. \cite{ev06} have studied the X-ray properties of the
cores of 22 low-redshift ($z < 0.1$) radio galaxies and find a strong
correlation between 1 keV X-ray and 5 GHz radio core powers for
low-absorption ($N_{\rm H} < 5 \times 10^{22}$ cm$^{-2}$) radio galaxies, a
sub-sample which includes most FR Is. For $\alpha_x = 0.9$ (their average
value) and $\alpha_r \sim 0$ (appropriate for radio cores) I get a mean
value $\langle \log(f_{\rm 0.5 - 2 keV}/S_{1.4GHz}) \rangle \sim -18$. One
might worry that for more distant sources it will be hard to resolve the
nuclear components. I then derived the typical X-ray-to-radio flux density
ratio for the FR Is in the AGN catalogue of \cite{pad97} by using total
X-ray and radio fluxes, obtaining $\langle \log(f_{\rm 0.5 - 2
  keV}/S_{1.4GHz}) \rangle \approx -19$. In the following I will then adopt
the intermediate value $\langle \log(f_{\rm 0.5 - 2 keV}/S_{1.4GHz})
\rangle \sim -18.5$, noting that even assuming the largest value would not
affect any of my conclusions.

Flux density ratios will be largely unaffected by evolution, since this is
broadly similar in the X-ray and radio band for SFG \citep{ran05} and AGN
\citep[e.g.,][]{wa05}. No information is available on the X-ray evolution
of FR Is, which in any case are weak X-ray emitters.

This approach has its limitations. The SFG used by
\cite{ran03} have $S_{\rm 1.4GHz} > 0.1$ Jy and the FR Is studied by
\cite{ev06} have $S_{\rm 5GHz} ({\rm core}) > 0.01$ Jy. The derived mean
X-ray-to-radio flux density ratios are then used to predict the X-ray
fluxes of radio sources in the $\mu$Jy and nanoJy regime, that is at $\ga
10^4$ times fainter flux density levels. Moreover, these values
are likely to be prone to selection effects, which I have tried to take
into account in the case of radio-quiet AGN. The latter could be less
relevant if these values are based on a strong, physically based
correlation, which is likely the case for SFG.

\subsubsection{Extrapolation from the VLA-CDFS Sample}\label{extrapolation}

An alternative, complementary procedure is to derive the mean X-ray flux
for a sample with a relatively faint radio flux density limit and then
shift it down to simulate a fainter sample. To this aim, I used the
VLA-CDFS sample for which \cite{pad09} (see also Padovani et al., in
preparation) provide a classification in SFG and AGN (radio-quiet and
radio-loud). Based on their results, most AGN of the radio-loud type are
expected to be low-luminosity radio-galaxies, that is FR Is. The flux
density limit of the VLA-CDFS survey is not constant but increases with
distance from the field centre from a minimum value of $42~\mu$Jy
\citep{kel08}. It then follows that low-flux density sources, which, on
average, have also the smallest X-ray fluxes, are underrepresented as
compared to a sample with a constant limit across the field. To correct for
this I have weighted each source by the inverse of the area associated to
its flux density. I then divided the mean values of the X-ray flux by 42 to
simulate a sample with $S_{\rm 1.4GHz} \ge 1~\mu$Jy. Since most SFG and
radio-loud AGN are undetected in the X-ray band, only upper limits are
available for these two classes. In an hypothetical sample with a radio
flux density limit of 1 $\mu$Jy SFG should have $\langle f_{\rm 0.5 - 2
  keV} \rangle < 5 \times 10^{-18}$ erg cm$^{-2}$ s$^{-1}$, FR Is should
have $\langle f_{\rm 0.5 - 2 keV} \rangle < 10^{-17}$ erg cm$^{-2}$
s$^{-1}$ and radio-quiet AGN should be characterised by 
$\langle f_{\rm 0.5 - 2 keV}
\rangle \sim 2 \times 10^{-17}$ erg cm$^{-2}$ s$^{-1}$. 

\begin{figure*}
\centering
\includegraphics[width=16.0cm]{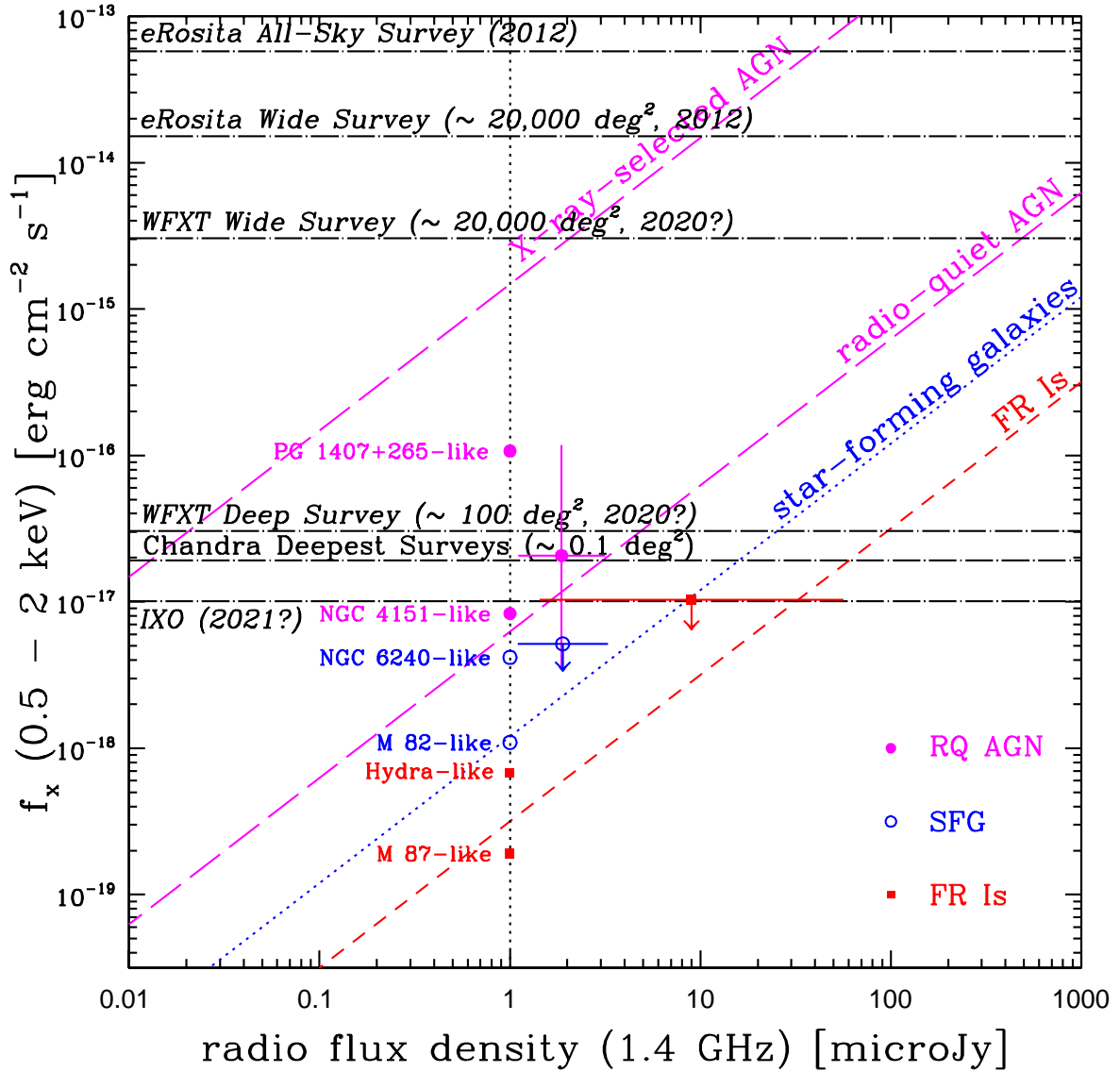}
\caption{0.5 -- 2 keV X-ray flux vs. the 1.4 GHz radio flux density for
  faint radio sources. The loci of X-ray selected and radio-selected,
  radio-quiet AGN (long-dashed lines), SFG (dotted line), and FR Is
  (short-dashed line) are indicated. The scaled X-ray fluxes of
  prototypical representatives of the three classes at $S_{\rm 1.4GHz} =
  1~\mu$Jy are also shown. Finally, the mean radio and X-ray flux values,
  or upper limits, for sources belonging to an hypothetical sample
  characterised by $S_{\rm 1.4GHz} \ge 1~\mu$Jy, as extrapolated from the
  VLA-CDFS sample, with error bars indicating the standard deviation, are
  also marked. The horizontal dot-dashed lines indicate the limits of (from
  top to bottom): the eRosita All-Sky and Wide Surveys, the WFXT Wide and
  Deep Surveys, Chandra's deepest surveys, and IXO. Survey areas and launch
  dates for future missions, or best guesses at the time of writing, are
  also shown. See text for more details.}
\label{fxfr} 
\end{figure*}
    
\subsubsection{X-ray fluxes of faint radio sources}\label{x-ray_fluxes}
    
Figure \ref{fxfr} plots 0.5 -- 2 keV flux vs. 1.4 GHz radio flux density
and shows the loci of X-ray selected and radio-selected, radio-quiet AGN,
SFG, and FR Is (Sect. \ref{ratios}). Note that radio-quiet AGN will span
the full range between the two dashed lines in Fig. \ref{fxfr}, with radio
(X-ray) selection favouring sources with low (high) X-ray-to-radio flux
density ratios. The position of these loci with respect to survey limits
determines the fraction of sources of a given class detected in one band
with counterparts in the other.
The figure shows also the expected X-ray fluxes for ``typical'' radio-quiet
AGN, SFG, and FR Is (from the NASA/IPAC Extragalactic Database [NED])
scaled to 1 $\mu$Jy.
The mean radio and X-ray flux values, or upper limits, for sources
belonging to an hypothetical sample characterised by $S_{\rm 1.4GHz}
\ge1~\mu$Jy (Sect. \ref{extrapolation}) are also shown. The fact that the
three estimates give consistent results is reassuring and shows that we can
predict reasonably well the X-ray fluxes of faint radio sources.

The horizontal dot-dashed line at $f_{\rm 0.5 - 2 keV} \sim 2 \times
10^{-17}$ erg cm$^{-2}$ s$^{-1}$ represents the deepest X-ray data
currently available, that is the Chandra Deep Field South 2 Ms Survey
\citep{luo08}, covering about 0.1 deg$^2$. The other horizontal lines
indicate, from top to bottom, the limiting fluxes for point sources for
surveys to be carried out with
eRosita\footnote{http://www.mpe.mpg.de/heg/www/Projects/EROSITA/\-main.html}
and the Wide Field X-ray Telescope\footnote{http://wfxt.pha.jhu.edu/}
(WFXT)\footnote{A WFXT-centric version of Fig. \ref{fxfr} has been
  presented by \cite{pad10}}. The faintest X-ray flux limit corresponds to
the International X-ray Observatory
(IXO)\footnote{http://ixo.gsfc.nasa.gov/}, which will provide the deepest
X-ray view on the Universe for quite some time. IXO will be an observatory
type mission, like Chandra, and therefore will not produce large area
surveys.

The main message of Fig. \ref{fxfr} is that even the most powerful X-ray
missions we are going to have for the next 20 years or so will only detect
the counterparts of radio-quiet AGN with radio flux densities down to
$\approx 1~\mu$Jy. The bulk of the $\mu$Jy population, which is most likely
to be made up of SFG, will have X-ray fluxes beyond even the reach of
IXO. Same, or possibly even worse, story for FR Is. The situation will
obviously be even more critical in the nanoJy regime, where very few radio
sources will have an X-ray counterpart in the foreseeable future.

On the positive side, basically all extragalactic sources in the eRosita
All-Sky and Wide Surveys and WFXT Wide Survey will have an SKA counterpart,
as they will have $S_{\rm 1.4GHz} > 1~\mu$Jy. This should help in the
identification work of, for example, the 10 million or so point sources
expected in the latter, by also providing very accurate positions. Radio
detection of the bulk of the AGN in the WFXT Deep Survey will require much
higher ($\approx 20$ nanoJy) sensitivities. This might be accomplished by
the SKA given also the small area of the survey ($\sim 100$ deg$^2$), under
the obvious condition that WFXT surveys are carried out in the southern
sky.

\subsection{Optical Band}

\subsubsection{Typical radio-to-optical flux density ratios}\label{ratios_opt}
 
At variance with the X-ray case, there is in general no apparent
correlation between radio and optical powers in extragalactic sources. This
is due to the fact that while the radio, far-infrared, and X-ray bands all
trace the star formation rate (SFR) \citep[e.g.,][]{ken98,ran03}, the
optical band does not, as young stars dominate the ultraviolet continuum.
Moreover, as mentioned in Sect. \ref{pop}, elliptical galaxies of similar
optical power can host radio sources differing by huge amounts in their
radio luminosity.

However, SFG and radio-quiet AGN can only reach a reasonably well-defined
value of the K-corrected radio-to-optical flux density ratio $R = \log
(S_{\rm 1.4GHz}/S_{\rm R_{\rm mag}})$, where $S_{\rm R_{\rm mag}}$ is the
R-band flux density \citep[e.g.,][]{mac99,pad09}.  Converting from the
values derived for the two classes in the V-band by \cite{pad09} and taking
the average one gets for the R-band a maximum value $R \approx 1.4$. Due to
K-correction effects (see below), {\it observed} $R$ values for SFG and
radio-quiet AGN can be $> 1.4$. It has to be noticed that, while all
``classical" radio-loud quasars have $R > 1.4$, this is not the case for
many radio-galaxies \citep[e.g.,][]{pad09}, which can extend to $R
<1.4$. Indeed, as mentioned above, low-power ellipticals will have lower
$R$ values than FR Is.

The absolute SFR in star-forming galaxies spans a very large range, from
$\sim 20~M_{\odot}$ yr$^{-1}$ in gas-rich spirals to $\sim 100~M_{\odot}$
yr$^{-1}$ in optically selected starburst galaxies and up to $\sim
1000~M_{\odot}$ yr$^{-1}$ in the most luminous IR starbursts \citep{ken98}.
Therefore, for the same optical magnitude ``normal'' spirals will be
characterised by lower radio emission than starbursts, and therefore will
have a smaller radio-to-optical flux density ratio.
To better quantify this I will define as a ``normal", non-starburst source
a galaxy with an SFR $< 10~M_{\odot}$ yr$^{-1}$, which is the maximum value
reached by local Uppsala Galaxy Catalogue (UGC) galaxies
\citep{jam04}. Converting this to a radio power following the calibration
of \cite{sarg09}, which agrees very well with that based on polycyclic
aromatic hydrocarbons (PAH), translates to $P_{\rm 1.4GHz} < 8.4 \times
10^{21}$ W Hz$^{-1}$. SFG in the VLA-CDFS sample below this value have $R
\la 0.3$ (and $z \la 0.2$). \cite{pad09} have also shown that $R$ decreases
with decreasing $P_{\rm 1.4GHz}$ for the spirals and irregulars in their
sample. From Fig. 1 of \cite{con89} (converting from his cosmology) I
derive $R \la 0.1$ for $P_{\rm 1.4GHz} \sim 8 \times 10^{21}$ W Hz$^{-1}$
for optically bright ($B_{\rm T} \le 12$) spirals and irregulars; lower
radio powers have smaller $R$ values.

Dwarf galaxies have also low flux density ratios and are in fact
underrepresented in radio samples \citep{van04}. From the mean values given
by \cite{le05} I get $R \approx -0.6$ (converting to my notation), while
Fig. 9 of \cite{leo08} shows that $R < -0.8$ for $M_B \sim -17$. Even BCDs,
which are the most star-forming amongst dwarfs, are characterised by $R
\approx -0.2$ \citep{hun05}.

In the optical band the K-correction is quite important. For example, at $z \sim
1$, the mean redshift of the VLA-CDFS sample, it reaches $\sim 1$ magnitude
for Sbc galaxies and $\sim 2$ magnitudes for ellipticals \citep[e.g.,]
[]{col80}. These values should be compared to the factor $\sim 1.2$ in flux
(equivalent to $\sim 0.2$ magnitudes) expected in the radio band (for
$\alpha_r \sim 0.7$). Moreover, absorption by the intergalactic medium will
also further decrease the optical flux. It then follows that the derived
magnitudes are very robust lower limits. As was the case for the X-ray
band, flux density ratios will be largely unaffected by evolution, since
this is broadly similar in the optical and radio band for SFG
\citep{ran05} and powerful AGN \citep{wa05}.  
For FR Is and low-power ellipticals, however, the evolutionary correction
in the optical band can be quite important \citep{po97}. This, coupled to
the almost absent radio evolution (Sect. \ref{pop}), could imply a decrease
in radio-to-optical flux density ratios for these classes at high
redshifts.

\subsubsection{Extrapolation from the VLA-CDFS Sample}\label{extrapolation_opt}

The best estimate of the optical magnitudes of faint radio sources is
obtained by scaling those of a sample with a relatively faint radio flux
density limit. Using the same procedure as for the X-ray band
(Sect. \ref{extrapolation}), I added to the $\langle R_{\rm mag} \rangle$
of the VLA-CDFS sources $2.5 \times \log 42$ to simulate a sample with
$S_{\rm 1.4GHz} \ge 1~\mu$Jy. Correcting for the non-uniform flux density
limit as done before (Sect. \ref{extrapolation}), SFG in this hypothetical
sample should have $\langle R_{\rm mag} \rangle \sim 26.3$, radio-quiet AGN
should have $\langle R_{\rm mag} \rangle \sim 28.2$ and radio-loud AGN
(mostly low-luminosity radio-galaxies) should have $\langle R_{\rm mag}
\rangle \sim 27.3$. Note that $\sim 85\%$ of the SFG in the VLA-CDFS sample
have radio powers above the ``normal'' spiral limit and so are mostly of
the starburst type.

\subsubsection{Optical magnitudes of faint radio sources}\label{opt_mag}

\begin{figure*}
\centering
\includegraphics[width=16.0cm]{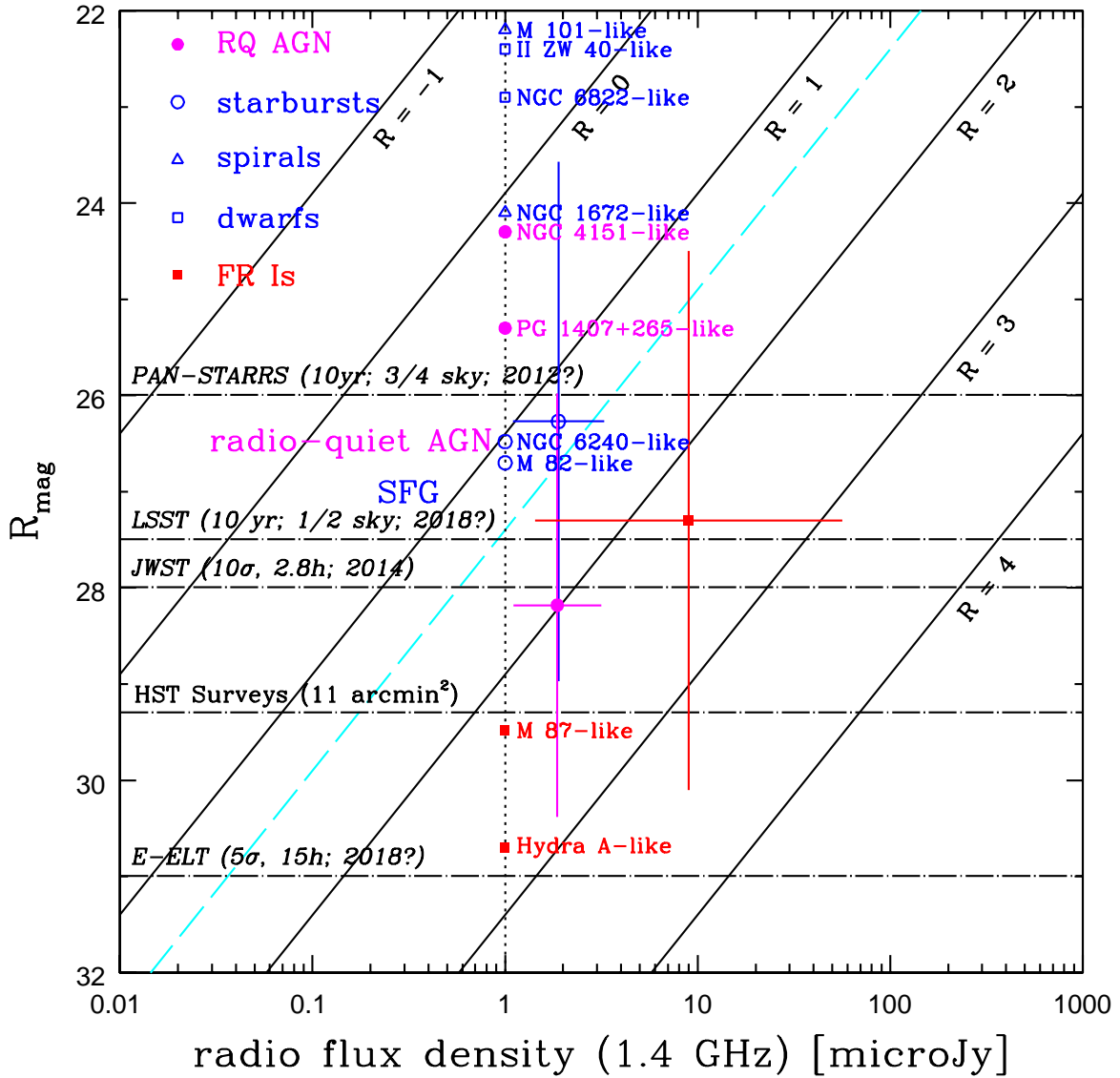}
\caption{$R_{\rm mag}$ vs. the 1.4 GHz radio flux density for faint radio
  sources. Diagonal lines represent different values of $R = \log (S_{\rm
    1.4GHz}/S_{\rm R_{\rm mag}})$, ranging from $-1$ (top) to 4 (bottom).
  The diagonal dashed line at $R=1.4$ indicates the maximum value for SFG
  and the approximate dividing line between radio-loud and radio-quiet AGN,
  with SFG and radio-quiet AGN expected to populate the top left part of
  the diagram.
  The scaled R magnitudes of prototypical representatives of the three
  classes at $S_{\rm 1.4GHz} = 1~\mu$Jy are also shown, with SFG split 
  into starbursts, spirals, and dwarfs. Finally, the mean radio and $R_{\rm
    mag}$ values for sources belonging to an hypothetical sample
  characterised by $S_{\rm 1.4GHz} \ge 1~\mu$Jy, as extrapolated from the
  VLA-CDFS sample, with error bars indicating the standard deviation, are
  also marked. The horizontal dot-dashed lines indicate the approximate
  point-source limits of (from top to bottom): PAN-STARRS, LSST, JWST, 
   the deepest Hubble Space Telescope surveys, and
  the E-ELT. 
  Survey areas (for LSST and PAN-STARSS), S/N ratios and exposure times
  (for the E-ELT and JWST) and operation/launch dates for future missions,
  or best guesses at the time of writing, are also shown. See text for more
  details.}
\label{Rfr} 
\end{figure*}
    
Figure \ref{Rfr} plots $R_{\rm mag}$ vs. 1.4 GHz radio flux density and
shows: 
1. the expected $R_{\rm mag}$ for ``typical'' radio-quiet AGN, SFG
(starbursts, spirals, and dwarfs), and FR Is (from NED) scaled to 1 $\mu$Jy;
2. the maximum value for SFG and the approximate dividing line between
radio-loud and radio-quiet AGN (diagonal dashed line; see
Sect. \ref{ratios_opt}). SFG and radio-quiet AGN are expected to populate
the top left part of the diagram; 3.  the mean radio and $R_{\rm mag}$
values for sources belonging to an hypothetical sample characterised by
$S_{\rm 1.4GHz} \ge 1~\mu$Jy (Sect.  \ref{extrapolation_opt}).

The three methods give consistent results, which is reassuring and shows
that we have a reasonable handle on the magnitudes of faint radio
sources. The scaled $R_{\rm mag}$ for radio-quiet AGN are fainter than
those derived for ``typical'' sources and are beyond the $R = 1.4$ line.
This is because these sources have the highest redshifts in the VLA-CDFS
sample, and therefore their observed magnitudes are more affected by
K-correction effects.

The horizontal dot-dashed line at $R_{\rm mag} \sim 29.3$ represents the
limit of the deepest optical data currently available, the Hubble UDF
\citep{bec06}, covering about 11 arcmin$^2$. The other horizontal lines
indicate, from top to bottom, the limiting magnitudes for the Panoramic
Survey Telescope and Rapid Response System
(PAN-STARRS)\footnote{http://pan-starrs.ifa.hawaii.edu/}, which from Hawaii
will survey about 3/4 of the sky down to $R_{\rm mag} \sim 26$ during 10
years of operation and the Large Synoptic Survey Telescope
(LSST)\footnote{http://www.lsst.org}, which will be located in Chile and
will provide a survey of about half the sky down to $R_{\rm mag} \sim 27.5$
during 10 years of operation. Further below there are the limiting
magnitudes for point sources for the James Webb Space Telescope
(JWST)\footnote{http://www.stsci.edu/jwst/} and for the European Extremely
Large Telescope\footnote{http://www.eso.org/sci/facilities/eelt/}
(E-ELT)\footnote{I am using the E-ELT as an example because it has the
  largest mirror (42 m in diameter) amongst the three very large telescopes
  being planned. The other two are the Thirty Meter Telescope (TMT;
  http://www.tmt.org/) and the Giant Magellan Telescope (GMT;
  http://www.gmto.org/).}  for the given signal-to-noise (S/N) ratios and
exposure times\footnote{The JWST limit comes from the JWST Web pages. The
  E-ELT limit was derived from the E-ELT Exposure Time Calculator (ETC)
  using the Laser-Tomography/Multi-Conjugate Adaptive Optics option.}. Both
these telescope will operate in observatory mode.
It is important to keep in mind that all of these limits are only
approximate, as none of these telescopes are in operation yet.


Given the progression towards lower $R$ values going from
(non-dwarf) starbursts, to spirals, and to BCDs and dwarf irregulars
(Sect. \ref{ratios_opt}), the typical magnitudes of faint radio
sources at a given flux density will depend critically on which of these
SFG sub-classes dominates the counts.

The contribution of ``normal'' spirals to faint radio number counts is not
well known. \cite{hop00}, by converting the optical LF to a radio LF
adopting a constant ratio $P_{\rm 1.4GHz}/L_{\rm B} = 1/3$, predict that
normal galaxies will outnumber starbursts at $S_{\rm 1.4GHz} \sim
1~\mu$Jy. Since $R$ is a function of radio power, this assumption is too
simplistic.
I have used a different approach based on the SFR division introduced above
(Sect. \ref{ratios_opt}). Namely, I have split the SFG LF of \cite{sad02}
into two at $P_{\rm 1.4GHz} = 8.4 \times 10^{21}$ W Hz$^{-1}$ and evaluated
the number counts assuming the evolution of \cite{hop04}. The result is
that ``normal'' spirals should be more numerous than starbursts for $S_{\rm
  1.4GHz} \la 5~\mu$Jy.

Are dwarf galaxies also going to be relevant at the flux densities of
interest here ($\ga 10$ nano Jy)? Although we do not have any information
on their radio LF, one can make an educated guess by converting their
optical LF. \cite{leo08} find that radio and optical power for Sb -- Sc
galaxies extending down to $M_{\rm B} \sim -16$ scale linearly; the scaling
factor is consistent with the mean values for the dwarf galaxies studied by
\cite{le05}, which reach $M_{\rm B} \sim -14$. I have then transformed the
dwarf LF of \cite{bla05} to a radio LF, which covers the range $10^{18} \la
P_{\rm 1.4GHz} \la 3 \times 10^{20}$ W Hz$^{-1}$. Assuming the same
evolution as SFG and integrating as done above to $z_{\rm max} = 3$, one
then finds that dwarfs outnumber non-dwarf SFG at $S_{\rm 1.4GHz} \la 20$
nanoJy. This result is quite robust towards extrapolations of the optical
LF to fainter magnitudes ($M_{\rm r} \sim -10$) with the same slope
($\alpha = -1.52$), while a steepening of the LF ($\alpha = -2.0$) would lead to
an increase in the flux density below which dwarfs dominate by a factor
$\sim 3$. Note also that any evolution smaller than the assumed one would
push further down the contribution of dwarf galaxies at a given flux
density. However, the range in radio power covered by this radio LF does
not extend to the values reached by some BCDs \citep[$P_{\rm 1.4GHz} > 5
\times 10^{20}$ W Hz$^{-1}$: e.g.,][]{hun05,sarg09}. I have then multiplied
the scaling factor by $10^{0.5}$ to take into account the somewhat larger
radio-to-optical flux density ratio of BCDs (see Sect. \ref{ratios_opt}),
with the result that dwarfs should outnumber non-dwarf SFG at $S_{\rm
  1.4GHz} \la 0.3~\mu$Jy. Since not all dwarfs are BCDs, dwarf galaxies
will become the most numerous constituents of the radio sky at flux
densities in the likely range 20 nanoJy $\la S_{\rm 1.4GHz} \la$ 300
nanoJy.

Fig. \ref{Rfr} shows that, down to $1~\mu$Jy, most starburst-like SFG and
radio-quiet AGN should be detected by the LSST, that is they will have a
counterpart in a large area survey. To be more specific, based on the
VLA-CDFS extrapolation, $\sim 66\%$ of starburst galaxies with $S_{\rm
  1.4GHz} \ge 1~\mu$Jy should have $R_{\rm mag} < 27.5$. This number
represents an upper limit because by scaling the VLA-CDFS magnitudes I have
assumed that the mean redshift is unchanged, while \cite{pad09} have found
a strong correlation between redshift and magnitude. Therefore,
K-correction effects will be larger and observed magnitudes will be
fainter. On the other hand, ``normal'' spirals should outnumber starbursts
for $S_{\rm 1.4GHz} \la 5~\mu$Jy. If even distant spirals are characterised
by rest-frame $R \la 0.1 -0.3$, as is the case for local sources
(Sect. \ref{ratios_opt}), then $\mu$Jy sources could be $\approx 2$
magnitudes brighter than expected in the case of starbursts and therefore
within reach of PAN-STARRS and certainly of the LSST. For fainter radio
samples optical magnitudes should get fainter, unless dwarf galaxies, with
their low $R$ values, become dominant, which would lead yet to another
``brightening''. For example, for $R \approx -0.5$ and $S_{\rm 1.4GHz}
\approx 100$ nanoJy, $R_{\rm mag} \approx 25$.

FR Is have on average larger $R$ values but also higher radio flux
densities, so their mean magnitude in an hypothetical $S_{\rm 1.4GHz} \ge
1~\mu$Jy sample is brighter than those for ``typical'' sources with $S_{\rm
  1.4GHz} = 1~\mu$Jy, although the $R$ values are similar. Low-power
ellipticals, however, will have smaller $R$ and therefore brighter
magnitudes.

Sources having $R_{\rm mag} > 27.5$ will obviously be within reach of JWST
and ELTs, which however will be covering a relatively small field of view
(up to a few arcmin$^2$). Depending on the actual relative fraction of
starbursts, spirals, and dwarfs, these observatories could be the main
(only?) facilities to secure optical counterparts of nanoJy radio sources.

To address the question of what fraction of sources in the LSST and
PAN-STARRS surveys will have an SKA counterpart requires a knowledge of the
radio properties of very faint optical sources, which at present we do not
possess. Nevertheless, one can make some educated guesses. The UDF number
counts of \cite{bec06} show that for $z_{\rm AB} \la 26$ the surface
density is $\sim 3 \times 10^5$ deg$^{-2}$, which implies, based on
Fig. \ref{flim}, that the majority of the objects beyond this limit are
star-forming systems. SFG in the VLA-CDFS sample have {\it observed} $R
\approx 1.3$, which means $S_{\rm 1.4GHz} \approx 3~\mu$Jy and $\approx 70$
nanoJy at $R_{\rm mag} \sim 26$ and $\sim 27.5$ respectively, that is
within reach of the SKA. Optical selection however will be biased towards
smaller $R$ and therefore fainter radio flux densities. More importantly,
spirals and dwarfs, which might be the most numerous sub-classes, with
their lower $R$ values will have fainter radio flux densities for a given
magnitude. As regards the small minority of radio-quiet AGN, they can reach
$R \sim -1.5$ in the AGN catalogue of \cite{pad97}, which would imply
extremely low radio flux densities even for PAN-STARSS. However,
\cite{whi07}, by stacking FIRST radio images for $\sim 40,000$ SDSS
quasars, have shown that $R$ increases with magnitude, which would imply
higher radio flux densities at faint magnitudes. In summary, the bulk of
LSST and PAN-STARRS sources {\it might not} have radio flux densities
within reach of the SKA, but at this point in time one cannot be more
specific.


\subsection{Mid-Infrared Band}

\subsubsection{Typical mid-infrared-to-radio flux density ratios}\label{ratios_IR}

The infrared and radio emission are strongly and linearly correlated in
SFG, defining what is known as the ``IR-radio relation'' \citep[e.
g.,][and references therein]{sar10}. The most recent determination of the
so-called (K-corrected) $q_{24}$ parameter, based on the COSMOS field and
using an IR/radio-selected sample to reduce selection biases, is $q_{24} =
\log(S_{24\mu m}/S_{\rm 1.4GHz}) =1.26\pm0.13$ \citep{sar10}. As mentioned
above (Sect. \ref{ratios}), radio selection will favour objects with larger
radio flux densities, and therefore smaller $q_{24}$, but given the
tightness of the relation the decrease is only $\sim 0.3$ (based on Tab. 3
of \cite{sar10}). BCDs also appear to have $q$ parameters globally
consistent with those of SFG \citep{hun05}, even though \cite{bel03} argues
that both the IR and radio luminosities of dwarf galaxies significantly
underestimate the SFR by similar amounts. Radio-quiet AGN have $q_{24}$
values similar to those of SFG, while radio-loud ones have much smaller
ones. For this band we cannot extrapolate from the VLA-CDFS sample since
most of the sources are undetected in the $24~\mu$m band in publicly
available catalogues.

Evolution in the mid-infrared and radio band is broadly similar for SFG
\citep{ran05} and AGN \citep[e.g.,][]{wee09}. Therefore, typical flux
density ratios will be largely unaffected by it. No information is
available on the mid-infrared evolution of FR Is, which in any case are
very weak IR emitters.

\subsubsection{Mid-Infrared flux densities of faint radio sources}\label{IR_fluxes}

\begin{figure*}
\centering
\includegraphics[width=16.0cm]{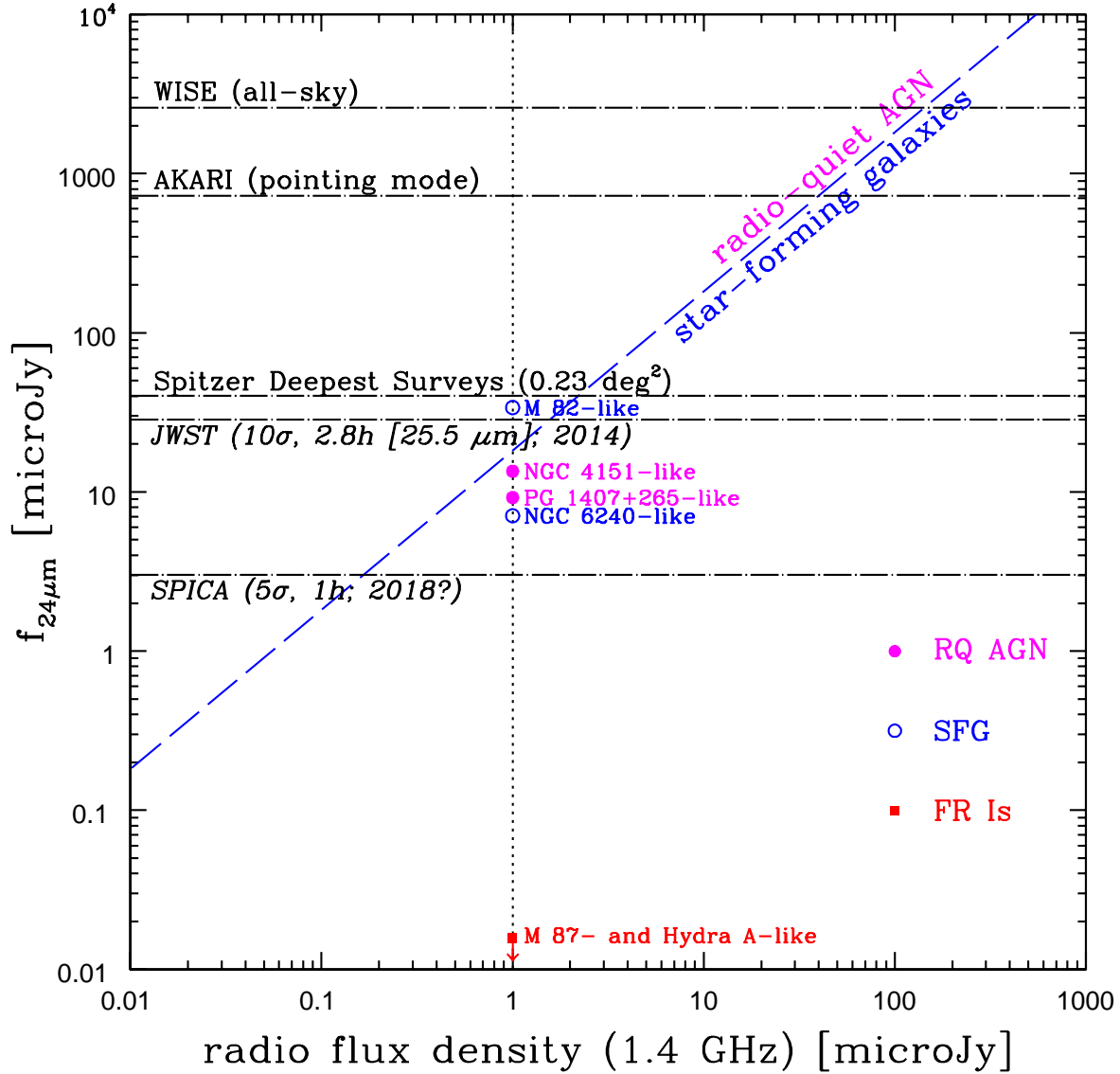}
\caption{$24 \mu$m flux density vs. the 1.4 GHz radio flux density for
  faint radio sources. The diagonal dashed line represents the locus of SFG
  and radio-quiet AGN based on the ``IR-radio relation".  The scaled IR
  flux densities of prototypical representatives of the three classes at
  $S_{\rm 1.4GHz} = 1~\mu$Jy are also shown, with FR Is being so faint as
  to be actually off the plot at $f_{\rm 24 \mu m} \sim 0.2 - 0.8$
  nanoJy. The horizontal dot-dashed lines indicate the approximate
  point-source limits of (from top to bottom): WISE, AKARI (pointing mode),
  the deepest Spitzer surveys, JWST, and SPICA. Launch dates for future
  missions, or best guesses at the time of writing, are also shown. See
  text for more details.}
\label{f24fr} 
\end{figure*}

Figure \ref{f24fr} plots the $24 \mu$m vs. 1.4 GHz radio flux density and
shows the expected $f_{\rm 24 \mu}$m for ``typical'' radio-quiet AGN and
SFG  
(from NED), scaled to 1 $\mu$Jy (the values for two ``typical'' dwarf galaxies 
are fully consistent with these and are not included for clarity)
and the locus
of SFG and radio-quiet AGN (Sect. \ref{ratios_IR}). The two methods give
perfectly consistent results. The two ``typical'' FR Is have $f_{\rm 24 \mu
  m} \sim 0.2 - 0.8$ nanoJy and are therefore way off the plot.

The two top horizontal dot-dashed lines represent the point-source limits
for the Wide-field Infrared Survey Explorer
(WISE)\footnote{http://wise.ssl.berkeley.edu/} and
AKARI\footnote{http://www.ir.isas.jaxa.jp/ASTRO-F/}. Both are performing
all-sky infrared surveys. However, I give the AKARI limit for pointing mode
as the survey limit for the band closest to the one of interest ($18~\mu$m)
is off the plot (120 mJy). The horizontal line at $f_{\rm 24 \mu m} \sim
40~\mu$Jy indicates the deepest mid-infrared data currently available, that
is the Spitzer observations of the GOODS/FIDEL field \citep{bet10},
covering about 0.23 deg$^2$. The bottom two horizontal lines denote the
limiting flux densities for point sources for JWST and the SPace Infrared
telescope for Cosmology and Astrophysics
(SPICA)\footnote{http://www.ir.isas.jaxa.jp/SPICA/} for the given S/N
ratios and exposure times. Both these telescope will operate in observatory
mode. These two limits are only approximate, as none of these telescopes
are in operation yet.

Figure \ref{f24fr} shows that most SFG and radio-quiet AGN should be
detected by JWST down to $\sim 1~\mu$Jy and by SPICA down to $\sim 100$
nanoJy. Both telescopes will be covering a relatively small field of view
(up to $\approx 15$ arcmin$^2$). Only a tiny fraction of $\mu$Jy and nanoJy
SFG and radio quiet AGN will have a counterpart in an all-sky infrared
surveys. As regards FR Is, they will be largely undetected even by SPICA.

On the other hand, all extragalactic sources in the WISE and AKARI all-sky
surveys will easily have an SKA counterpart, as they will be characterised by 
$S_{\rm 1.4GHz} \ga 100~\mu$Jy.

\section{Redshifts of faint radio sources}\label{redshifts}

\begin{figure*}
\centering
\includegraphics[width=16.0cm]{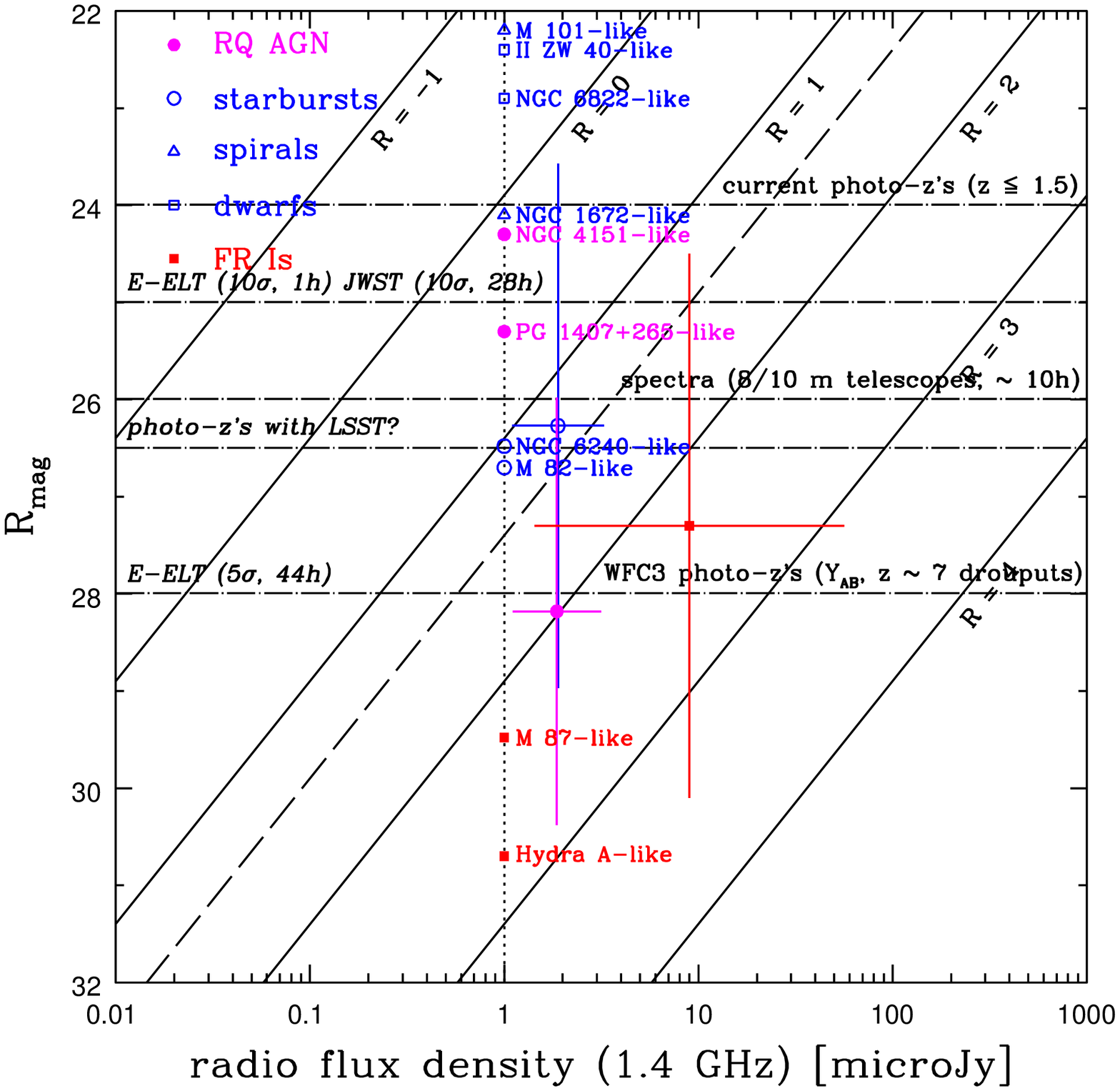}
\caption{$R_{\rm mag}$ vs. the 1.4 GHz radio flux density for faint radio
  sources. Diagonal lines represent different values of $R = \log (S_{\rm
    1.4GHz}/S_{\rm R_{\rm mag}})$, ranging from $-1$ (top) to 4
  (bottom). The diagonal dashed line at $R=1.4$ indicates the maximum value
  for SFG and the approximate dividing line between radio-loud and
  radio-quiet AGN, with SFG and radio-quiet AGN expected to populate the
  top left part of the diagram. 
  The scaled R magnitudes of prototypical representatives of the three
  classes at $S_{\rm 1.4GHz} = 1~\mu$Jy are also shown, with SFG split
  into starbursts, spirals, and dwarfs. Finally, the mean radio and $R_{\rm
    mag}$ values for sources belonging to an hypothetical sample
  characterised by $S_{\rm 1.4GHz} \ge 1~\mu$Jy, as extrapolated from the
  VLA-CDFS sample, are also marked. The horizontal dot-dashed lines
  indicate the approximate limiting magnitudes to derive photometric and
  spectroscopic redshifts. From top to bottom: the current limit for
  photometric redshifts, the JWST and an E-ELT limit (for the given S/N
  ratios and exposure times), the faintest magnitude for which a
  spectroscopic redshift can be obtained with 8/10 m telescopes, a possible
  limit for photometric redshifts with the LSST, a fainter E-ELT limit, and
  the limit of photometric redshifts for $z \sim 7$ drop-outs with the Wide
  Field Camera 3 (WFC3) on board HST. See text for more details.}
 \label{spectra} 
\end{figure*}

A vital component in the identification of astronomical sources is
redshift, which allows powers to be estimated and LFs to be derived. Figure
\ref{spectra} plots $R_{\rm mag}$ vs. 1.4 GHz radio flux density but, at
variance from Fig. \ref{Rfr}, the marked limits refer to spectroscopic or
photometric redshifts.

The top horizontal dot-dashed line represents the current limiting
magnitude for ``classical'' photometric redshifts, which reach $R_{\rm mag}
\approx 24$ and $z \approx 1.5$ \citep[e.g.,][]{mai08}; photometric
redshifts for special classes of sources, e.g., $z \sim 7$ drop-outs can be
derived with HST down to $Y_{\rm AB} \approx 28$ \citep{bou10}. Long
exposures ($\sim 10$h) with 8/10 m telescopes can secure spectroscopic
redshifts {\it in the case of strong emission lines} down to $R_{\rm mag}
\approx 26$, so this value represents a hard limit. Finally, indicative
limiting magnitudes for obtaining spectra for point sources with the E-ELT
and JWST are also shown\footnote{The JWST limit refers to a resolution of
  1,000 and $\lambda = 1~\mu$m and comes from the JWST Web pages. The E-ELT
  limits apply to a resolution of 1,000, the $R$ band, and was derived from
  the E-ELT ETC using the Laser-Tomography/Multi-Conjugate Adaptive Optics
  option. Since the choice of instruments for the E-ELT has not been
  finalised yet, these limits are {\it very} preliminary. Note that in both
  cases the magnitudes refer to the continuum: the presence of emission
  lines will obviously lead to shorter exposure times.}.

Fig. \ref{spectra} shows that, down to $1~\mu$Jy, only approximately half
of the starburst-like SFG will be within reach of 8/10 m telescopes in
terms of obtaining a spectroscopic redshift. To be more specific, based on
the VLA-CDFS extrapolation, $\sim 50\%$ of starburst galaxies with $S_{\rm
  1.4GHz} \ge 1~\mu$Jy should have $R_{\rm mag} > 26$ and $\sim 60\%$ will
be above $R_{\rm mag} = 25$, probably a more realistic limit for a
spectrum. As discussed above (Sect. \ref{opt_mag}), these fractions are
robust upper limits. On the other hand, spiral, and maybe dwarf galaxies,
should be more common than starbursts at $\mu$Jy and nano-Jy levels,
respectively, which would lead to brighter magnitudes, thereby making the
job of deriving a redshift easier. Future facilities like the E-ELT and
JWST will obviously allow the determination of spectroscopic redshifts for
fainter ($R_{\rm mag} > 26$) objects, albeit with relatively long
exposures. It might be also reasonable to suppose that the LSST will reach
the required accuracy to determine photometric redshifts for sources one
magnitude brighter than its 10 yr survey limit, thereby reaching $R_{\rm
  mag} \approx 26.5$. However, both LSST and PAN-STARRS will work with
photometric systems with $\sim 6$ optical broadband filters, similar to
those traditionally used in astronomy, which means that they will be more
prone to color-redshift degeneracies than systems using a larger number of
narrower filters \citep{ben09}.  Since the reddest filter is going to be in
the $y$-band ($\sim 1.05~\mu$m), this will also limit the maximum
achievable redshift.

Note that the magnitude limits of the current deepest large-area
spectroscopic surveys are $r \le 17.8$ (galaxies) and $i \le 19.1$
(quasars) for the SDSS\footnote{http://www.sdss.org/dr7/} and $b_J \le
20.85$ for the 2dF QSO redshift survey\footnote{http://www.2dfquasar.org/}.
 
In summary, many of the $S_{\rm 1.4GHz} \ge 1~\mu$Jy sources might be too
faint for 8/10 m telescopes to be able to provide a redshift and the
situation might get worse at fainter flux densities, unless dwarf galaxies
take over. This means that JWST and the ELTs might be the main facilities
to secure redshifts of $\mu$Jy radio sources. But even they could have
problems in the nanoJy regime.

\section{Discussion}\label{disc}

The simple approach employed in this paper suggests that (non-dwarf) SFG
play a very important role in the sub-$\mu$Jy sky, something which had been
already realised by many authors. What is new here is that also low-power
($P_{\rm 1.4GHz} < 10^{20}$ W Hz$^{-1}$) ellipticals but especially dwarf
galaxies are very relevant, with the latter very likely being the most
important constituent of the faint radio sky. The arguments I have used to
reach this conclusions appear quite robust. Indeed, by limiting the
integration to $z_{\rm max} = 3$ for lack of information, the derived
values on the surface densities of these classes are lower limits.
\cite{chri09} have recently used a new maximum likelihood method to derive
a LF down to $M_{\rm r} \approx -14$, which displays a faint end slope
$\alpha = -1.62$ steeper than the LF of \cite{bla05} I have used for my
calculations (which has $\alpha = -1.52$). The number (and therefore
surface) density of dwarf galaxies is then probably even higher than I have
assumed.

The radio, far-infrared, and X-ray bands all trace star formation, which
implies relatively tight relationships between these three bands. This
means that we can reasonably predict the mid-infrared and X-ray properties
of SFG with faint radio flux densities. The situation is different in the
optical band, which is dominated by emission from evolved stars. Since
``normal'' spirals and dwarf galaxies have radio-to-optical flux density
ratios smaller than starbursts because of their lower star formation rate,
they will be brighter in the optical band at a given radio flux density
ratio. The optical properties of faint radio sources will then depend on
the population mix of starbursts, spirals, and dwarf galaxies, which at
present cannot be determined very accurately.
Given our ignorance of the radio LF and evolution of dwarf galaxies, in
fact, it is not straightforward to predict at which flux densities they
will become the dominant population. By converting the optical LF to a
radio LF, under the assumption (not inconsistent with the available data)
of a linear proportionality between the two powers and of an evolution
similar to that of SFG, dwarfs should outnumber non-dwarf SFG in the 20 --
300 nanoJy range. However, this cannot be the whole story, as the estimated
radio LF reaches only $P_{\rm 1.4GHz} \sim 10^{21}$ W Hz$^{-1}$, while
there are BCDs with $P_{\rm 1.4GHz}$ up to $\approx 10^{22}$ W Hz$^{-1}$
\citep{sarg09}. There has then to be a high-power component of the LF, which
will translate into a higher number of sources at high flux densities,
which at present cannot be quantified.

Can the two ``new'' populations be relevant also for the radio background?
\cite{sin10} have concluded that the main background contributors have to
be faint (sub-$\mu$Jy) radio sources and suggested these to be ordinary
star-forming galaxies at $z > 1$ characterised by an evolving radio
far-infrared correlation, which increases toward the radio loud with
redshift (but see \cite{sar10} for evidence that the local radio
far-infrared relation still holds up to $z \sim 5$). \cite{mas10} (see also
Padovani et al., in preparation) have shown that population synthesis
models of the radio sky, which currently involve only ``classical'' radio
sources, fail to explain by a factor of a few (see their Fig. 9) the very
recent background measurements of \cite{fix10}. Since background emission
from a population of objects is equal to $\int S N(S) dS$ (where $N(S)$ are
the differential counts), which is a combination of surface and flux
density, the right mix of numbers and fluxes are needed to provide a
relevant contribution. Therefore, while low-power ellipticals, which are
both faint and not that numerous, are most likely unimportant, for dwarf
galaxies the situation is less obvious. On one hand, even with the largest
scaling factor used in Sect. \ref{opt_mag}, dwarf galaxies account for only
$\approx 8\%$ of the sub-mJy background. But on the other hand, the radio
LF is not known and, as discussed above, the radio number counts of dwarf
galaxies could well be higher.

The determination of the surface densities for all classes of radio sources
allows us also to determine the {\it intrinsic} ratio between radio-loud
and radio-quiet AGN. The surface density of SSRQs and FSRQs combined is
$\sim 6.3$ deg$^{-2}$, while that of radio-quiet, broad-lined AGN is
$\approx 5 \times 10^3$ deg$^{-2}$, since unabsorbed AGN are supposed to
make up $\sim 1/5$ of the total \citep{gil07}. This translates to a
radio-loud fraction in broad-lined AGN
$\approx 0.1\%$. If one considers both broad- and
narrow-lined sources, which means including also FR IIs in the radio-loud
class and all radio-quiet AGN,
one gets a similar fraction $\approx 0.2\%$. These numbers are much
smaller than the oft-quoted value of $\approx 10 - 20\%$. However, it has
been known for some time that the radio-loud fraction drops with decreasing
optical luminosity \citep{pad93}. Very recently, \cite{ji07} have shown
that the fraction of radio-loud quasars at $z=0.5$ declines from 24\% to
6\% as luminosity decreases from $M_{2500} = -26$ to $-22$, while at
$M_{2500} = -26$ this fractions declines from 24\% to 4\% as redshift
increases from 0.5 to 3. Both these results suggest that the intrinsic,
global ratio has to be $\la 5\%$. But if the radio-loud fraction were this
large, the radio-loud quasar surface density should be $\approx 250$
deg$^{-2}$, that is $\sim 40$ times larger than estimated here and also
$\sim 5$ larger than that of FR IIs, which are supposed to be the parent
population of both SSRQs and FSRQs. Scaling up the FR II surface density as
well to avoid this paradox, one would end up having more FR IIs than FR Is,
which would not make any sense at all, since FR IIs are more powerful. In
short, the surface densities derived in this paper point to an AGN
radio-loud fraction much smaller than normally derived. This difference is
likely to be ascribed to two factors: a) the usually quoted value has been
obtained for bright, optically selected samples \citep{kel89}, which
include radio-loud quasars with their optical flux boosted by relativistic
beaming \citep{gol99}, which artificially increases their fraction; b)
radio-loud AGN are on average more powerful than radio-quiet ones
\citep[e.g.,][]{zam08}. When one reaches the very faint end of the optical
LF only radio-quiet AGN will be present and therefore the integrated
radio-loud fraction will be quite small. In other words, the usually quoted
value refers to the bright part of the LF and, when integrated over the
full range of powers, the resulting radio-loud fraction is much
smaller. The second factor is likely to be the most relevant one.

The issue of source confusion goes beyond the scope of this paper. I will
just point out that \cite{win08} have shown that ultra-deep radio and
optical surveys may slowly approach the natural confusion limit, where
objects start to overlap with their neighbours due to their finite sizes
and not because of the finite instrumental resolution, which causes the
instrumental confusion limit. Based on the model of \cite{win03}, confusion
will not be a problem for the SKA if source sizes are $\la 1.6"$ at $S_{\rm
  1.4GHz} = 1~\mu$Jy (or $\la 0.25"$ at $S_{\rm 1.4GHz} = 10$ nanoJy) and
the instrumental resolution is commensurate \citep{win08}.  
However, since \cite{win03} did not include radio-quiet AGN, low-power
ellipticals, or dwarf galaxies, confusion limits are bound to be higher.
Indeed, \cite{ow08}, based on the relatively flat number counts derived
from their deep radio data, have suggested that the natural confusion limit
may be reached near $1~\mu$Jy.

\section{Conclusions}\label{conc}

I have used a simple method to work out the source population of the
sub-$\mu$Jy sky, which relies on estimating the smallest flux density and
the largest surface density for all classes of radio sources. These two
parameters can be derived quite robustly from the local minimum radio power
and number of sources per unit volume, maximum redshift, and luminosity and
density evolution. I have then estimated for the first time the X-ray,
optical, and mid-infrared fluxes these faint radio sources are likely to
have by using prototypical sources, typical flux density ratios, and
extrapolations from the VLA-CDFS sample. Prognosticating these
multi-wavelength properties is extremely relevant for the identification of
$\mu$Jy and nanoJy radio sources and to maximise the synergy between the
SKA and its pathfinders with future missions in bands other than the
radio. My main results can be summarised as follows:

\begin{enumerate}
 
 \renewcommand{\theenumi}{(\arabic{enumi})} 
 
\item The sub-$\mu$Jy sky should consist of radio-quiet AGN ($\approx 2
  \times 10^4$ deg$^{-2}$) and star-forming galaxies ($\approx 3 \times
  10^6$ deg$^{-2}$), both of which should get to $S_{\rm 1.4GHz} \approx
  0.1 - 1$ nanoJy. In agreement with previous studies, I find that
  classical, powerful radio sources, that is radio quasars and FR IIs, do
  not make it to sub-$\mu$Jy flux densities and reach $S_{\rm 1.4GHz}
  \approx 0.1$ mJy, with BL Lacs and FR Is getting to the $\approx 1 -
  10~\mu$Jy level.

\item The {\it intrinsic} fraction of radio-loud AGN, integrated over the
  whole bolometric luminosity function, is $\approx 0.1 - 0.2\%$, that is
  about two orders of magnitude smaller than the oft-quoted value of
  $\approx 10 - 20\%$.  The latter value refers only to the bright part of
  the optical luminosity function and is also biased because of the likely
  flux boosting in radio-loud quasars due to relativistic beaming.
     
\item Two ``new'' populations, which have not been considered previously,
  appear to be very relevant: low-power ($P_{\rm 1.4GHz} < 10^{20}$ W
  Hz$^{-1}$) ellipticals and dwarf galaxies. Using the available (scanty)
  information, the former should reach similar flux and surface densities
  as radio-quiet AGN, while the latter could easily be the most numerous
  component of the faint radio sky ($\ga 5 \times 10^6$ deg$^{-2}$), with
  flux densities as low as $\approx 1$ nanoJy. Since most galaxies at the
  faint end of the LF are blue, most dwarfs in the Universe (and therefore
  most radio sources) should be of the star-forming type.  While low-power
  ellipticals are most likely unimportant contributors to the radio
  background, the verdict is still open for dwarf galaxies. This issue is
  quite important also in light of the very recent background measurements
  reported by the ARCADE 2 collaboration \citep{fix10}.
  
\item The bulk of the $\mu$Jy population, which is then most probably going
  to be made up of star-forming galaxies, is likely to have X-ray fluxes
  beyond the reach of all currently planned X-ray missions, including
  IXO. The same applies to FR Is. Even IXO will only detect radio-quiet AGN
  with radio flux densities $\ga 1~\mu$Jy. The situation will obviously be
  worse in the nanoJy regime, where very few radio sources will have an
  X-ray counterpart in the foreseeable future. On the other hand, basically
  all extragalactic sources in the eRosita All-Sky and Wide Surveys and
  WFXT Wide Survey will have an SKA counterpart, as they will be
  characterised by $S_{\rm 1.4GHz} > 1~\mu$Jy.  This will help in the
  identification of, for example, the 10 million or so point sources
  expected in the latter, by also providing very accurate positions.

\item In the mid-infrared most star-forming galaxies and radio-quiet AGN
  should be detected by JWST down to radio flux densities $\sim 1~\mu$Jy
  and by SPICA down to $\sim 100$ nanoJy, while FR Is will be largely
  undetected even by SPICA.  Both telescopes will however be covering a
  relatively small field of view. Only a minute fraction of $\mu$Jy and
  nanoJy star-forming galaxies and radio quiet AGN will have a counterpart
  in current all-sky infrared surveys (WISE and AKARI). On the positive
  side, all extragalactic sources in the WISE and AKARI all-sky surveys
  will easily have an SKA counterpart, as they will be characterised by
  $S_{\rm 1.4GHz} \ga 100~\mu$Jy.

\item Since the radio, far-infrared, and X-ray emission all trace star
  formation, this implies relatively tight relationships between these
  three bands, which in turn means that we can reasonably predict the
  mid-infrared and X-ray properties of star-forming galaxies with faint
  radio flux densities. The situation is different in the optical band,
  where evolved stars dominate. Objects characterised by higher star
  formation rates will have larger radio-to-optical flux density ratios,
  which means fainter magnitudes for a given radio flux density. The
  typical magnitudes of the optical counterparts of faint radio sources
  depend then on which type of star-forming galaxy (starburst, spiral, or
  dwarf) will be predominant. Moreover, in the optical band K-correction
  and intergalactic absorption effects are important and will result in
  fainter than expected magnitudes. Assuming a maximum value of star
  formation rate for ``normal'' spirals, which translates to a maximum
  radio power, I estimate that these sources should outnumber starbursts
  for $S_{\rm 1.4GHz} \la 5~\mu$Jy. If even distant spirals are
  characterised by relatively low rest-frame radio-to-optical flux density
  ratios, then most $\mu$Jy sources should be detected by PAN-STARRS and
  certainly by the LSST. At fainter radio flux densities optical magnitudes
  should also get fainter, unless dwarf galaxies, with their lower
  radio-to-optical flux density ratios, become dominant. Depending on the
  relative fraction of starbursts, spirals, and dwarfs, JWST and especially
  the ELTs could be the main, or perhaps even the only, facilities capable
  of securing optical counterparts of nanoJy radio sources. On the other
  hand, and for the same reasons discussed above, the bulk of PAN-STARRS
  and LSST sources might not have radio flux densities within reach of the
  SKA.

\item As regards redshifts, the same complications described above apply,
  since the optical band is still involved.
  Within the same uncertainties, then, many of the sources with $S_{\rm
    1.4GHz} \ge 1~\mu$Jy might be too faint for 8/10 m telescopes to be
  able to provide a redshift determination and the situation might get
  worse at fainter flux densities, unless dwarf galaxies are
  prevalent. This means that JWST and particularly the ELTs might be the
  primary facilities to secure redshifts of $\mu$Jy radio sources. But even
  they could have problems in the nanoJy regime.

\end{enumerate}

In summary, the SKA and its pathfinders will have a huge impact on a number
of open problems in extragalactic astronomy. Apart from the ``obvious''
study of star-forming galaxies in their various incarnations and
``classical'' radio sources, these include also less evident ones, ranging
from what makes a galaxy radio-loud (through the study of low-power
ellipticals), to why most AGN are radio-quiet (by selecting large samples
independent of obscuration), to the incidence and evolution of dwarf
galaxies (by providing a ``cleaner'' radio selection, which might by-pass
the surface brightness problems of optical samples), to the resolution of
the radio background.

Identifying faint radio sources, however, will not be easy. On large areas
of the sky the SKA will be quite alone in the multi-wavelength arena, with
the likely exception of the optical band and even there probably only down
to $\approx 1~\mu$Jy. SPICA, JWST, and especially the ELTs will be a match
for the SKA but only on small areas and above $0.1 - 1~\mu$Jy. At fainter
flux densities one might have to resort to ``radio only" information, that
is HI redshifts, size, morphology, spectral index, etc., although I think
this will not be sufficient. On the bright side, most sources from
currently planned all-sky surveys, with the likely exception of the optical
ones, will have an SKA counterpart.

\section*{Acknowledgments}

The idea for this paper came to me while preparing a talk for the SKA 2010
meeting held in Manchester, UK, in March 2010. My thanks to the organisers
of that meeting for inviting me.  
I also thank Michael Hilker, Ken Kellermann, Jochen Liske, Vincenzo
Mainieri, Nicola Menci, Steffen Mieske, and Piero Rosati for useful
discussions, and the anonymous referee for his/her suggestions.
This research has made use of the
NASA/IPAC Extragalactic Database (NED) which is operated by the Jet
Propulsion Laboratory, California Institute of Technology, under contract
with the National Aeronautics and Space Administration and of NASA's
Astrophysics Data System (ADS) Bibliographic Services.

\label{lastpage}

\end{document}